# Optimizing Transmission FLASH Radiotherapy for Large-Field Post-Mastectomy Breast Treatment

Ahmal Jawad Zafar[1], Sunil William Dutta, Matthew Joseph Case, Zachary Diamond, Duncan Bohannon, Reshma Jagsi, Xiaofeng Yang, and Jun Zhou[1]*

[1]Department of Radiation Oncology and Winship Cancer Institute,
Emory University, Atlanta, GA, 30322, USA
*Email: jun.zhou@emory.edu

**Abstract**

**Purpose:** We investigated the effect of scanning speed, beam configuration, and dose-rate modeling on the FLASH effect in post-mastectomy proton transmission beams (TB) planning and evaluated the potential of spot scanning path optimization for enhancing the FLASH effect.

**Methods**: Five left-sided post-mastectomy breast patients (32 Gy/5 fractions) were replanned with single-energy (249 MeV) tangential TBs supplemented by a clinical an en face background beam. FLASH evaluation employed two models: Krieger's FLASH effectiveness model (FEM) and Folkerts' average dose rate (ADR) framework. Plans were delivered under conventional Pencil Beam scanning, split-field, and optimized spot sequences (using Genetic Algorithm (GA)), with vertical scan speeds varied from 10 to 20 mm/ms. FLASH effect in normal tissues was quantified by the percentage of voxels meeting the threshold (≥4 Gy at ≥40 Gy/s), and the dose adjustment factor of 0.67 was used once a voxel met FLASH criteria.

**Results**: The FLASH effect showed high sensitivity to scanning patterns and model selection. Increasing vertical scan speed from 10 to 20 mm/ms increased the FLASH in CTV by 22% (ADR) and from 12% (FEM), while in skin it rose from 41.4% to 58.8% (ADR) and 8.4% to 13.1% (FEM). Split-field delivery improved the temporal distance between the vertical columns of the spot scanning pattern, yielding a superior FLASH effect, which is up to a 9.2 Gy reduction in CTV Dmean with the ADR model. GA-based optimization shortened scan time and provided FLASH comparable to split-field delivery, with a CTV Dmean reduction of 7.87 Gy (ADR GA), with skin Dmean reductions of 2–3 Gy.

**Conclusion**: This study demonstrates that FLASH outcomes are highly sensitive to scanning trajectory, scan speed, and model selection. Beyond these parameters, optimizing spot delivery using a path minimizer, such as GA, can further improve the dose rate distribution in healthy voxels across all scenarios.

## 1. Introduction:

Radiation therapy is frequently advised in the treatment of breast cancer to improve local control and, in certain cases, improve overall survival. Surveillance, Epidemiology, and End Results (SEER) data estimates 316,950 new breast cancer diagnoses in the USA in 2025[1]. Some estimates have shown radiation is advised in up to 70% of cases[2]. The SEER estimated relative 5-year survival of breast cancer patients of 91.7% from 2015-2021 is favorable compared with the prognosis for many other types of cancers, highlighting the importance of reducing long-term toxicities associated with treatments[1].

Proton radiation can achieve similar coverage to photon therapy, with significantly reduced dose to organs at risk (OARs)[3]. Historically, one of the leading causes of post-photon radiotherapy mortality is a cardiovascular event attributed to coronary artery lesions[4-6], directly related to the radiation dose received by heart[5,7]. Proton radiation has been observed to deposit a reduced mean heart dose compared to the conventional RT, often with mean heart doses of <1 Gy[8-10]. Specific to left breast cancer cases, proton treatment deposits a lesser dose in the left anterior descending (LAD) and heart collectively[11-13].

In conventional proton therapy (PT), the Bragg peak is typically placed with the target volume using en face beams to maximize dose conformity. However, an alternative and less explored irradiation approach is the use of tangential transmission beams (TB), where the proximal portion delivers the dose of the beam, and Bragg peaks are placed outside the body[14-16]. While tangential beams are not used clinically in PT as this beam angle would negate the potential heart and lung sparing, this approach has some distinct advantages compared to en face beams, including enhanced robustness, reduced density-related uncertainties, and the ability to deliver ultra-high dose rates (UHDRs) with a single energy beam configuration[17]. Due to these characteristics, the TB approach is particularly well-suited for FLASH radiotherapy (FLASH-RT), which is defined by UHDRs exceeding 40 Gy/s and has been associated with enhanced tumor control and superior sparing of OARs. Although achieving UHDR in tangential photon beams would be technically challenging for standard LINAC beams, standard proton machines are more readily modified to achieve this effect.

Although the experimental FLASH data are still sparse, the preclinical evidence suggests that OARs lying within or near the target experience substantially less damage than observed with conventional dose rates[18,19]. Several preclinical studies have demonstrated the reduction in toxicity to healthy tissues while maintaining a comparable plan quality to conventional RT[18-21]. Van Marlen et al demonstrated that a single ultra–high dose-rate (UHDR) for 250 MeV proton TB with optimized beam splitting resulted in FLASH dose rates to the whole breast[22]. More recent studies have also demonstrated the use single energy beam with modularized pin-ridge filters (pRFs) to study the feasibility and adaptability of UHDR proton FLASH planning[23-25].

In pursuit of finding a biologically accurate definition of dose and dose rate, studies have used various models[15,26-28]. Since the exact mechanism behind FLASH is not yet fully understood, FLASH studies are presently divided into two types of dose rate definitions: average dose rate (ADR) and instantaneous dose rate. Although ADR metrics have been extensively used in FLASH evaluation, there are fundamental limitations in the models as they do not account for the scan time between the spots[28]. Folkerts et al. have published a PBS dose rate model that overcomes this limitation of scan time[26]. Although this voxel-wise average dose rate model yields a binary manner of FLASH and no-FLASH voxels, studies such as that by Vozenin et al. have shown the empirical evidence of the average dose rate to be the dominant variable to cause the FLASH effect[29]. Similarly, Petersson et al. highlighted the relevance of the oxygen depletion by total exposure over total time[30]. These studies have emphasized the importance of the average dose rate in FLASH studies.

The efficacy of FLASH RT is intricately connected to the dose rate. Several studies have explored PBS-specific dose rate optimization strategies, highlighting the critical role of the spot-delivery pattern[31,32]. The application of FLASH-RT by using TB on the chest wall presents a uniquely significant challenge due to the large target volume combined with the proximity of dosimetrically relevant OARs. Through dosimetric analyses of 5 post-mastectomy patients using UHDR TB-PT, this study aims to analyze the key parameters influencing PBS spot scanning patterns, compare the results of a previously reported and newly proposed FLASH-effect model applied to our data, and ultimately evaluate how effectively UHDR TB-PT can achieve adequate target coverage while preserving normal tissue sparing under varying scanning patterns and FLASH-effect models.

## 2. Methods and Materials:

### 2.1. Treatment planning:

A targeted delivery plan was designed and optimized for five post-mastectomy patients using RayStation. Typically, such optimization involves a pencil beam scanning (PBS) process where each spot's position and weight are optimized in accordance with clinical requirements. However, in the context of FLASH TB delivery using high-energy protons (e.g.,

249 MeV at our proton center), spot placement was performed manually to ensure sufficient coverage of the post-mastectomy chest wall target volume. Figure 1(d) shows a beam's eye view (BEV) spot map with uniformly spaced 249 MeV spots for a tangential TB with a gantry angle of approximately 310.8°. In this configuration, inter-spot spacing is a key parameter influencing both dose conformity and FLASH dose rate. While reduced spot spacing can enhance dose conformity, it may adversely affect dose rates, particularly in the treatment of large target volumes such as the post-mastectomy chest wall. Conversely, excessive spacing may result in dose inhomogeneities. In this study, a fixed inter-spot spacing of 6 mm was employed, which provided a reasonably homogeneous dose distribution across the target volume, as evidenced in Figure 1(a).

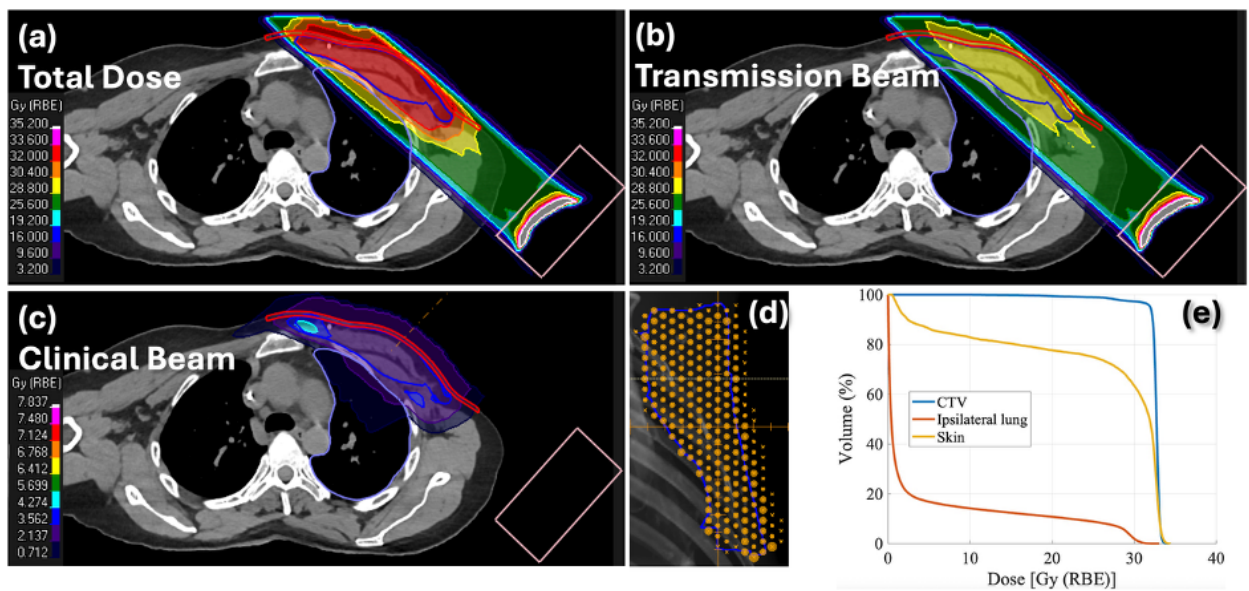

*Figure 1. Comparison of total dose, beam configurations, and dose–volume histograms (DVHs) for a representative post-mastectomy proton therapy case, Case 2 of this study. (a) Total dose distribution resulting from the combination of clinical and transmission beam (TB). (b) The dose contribution from the single-energy (249 MeV) TB, used in delivering the FLASH dose. (c) Dose distribution from the clinical beam as a background beamset, contributing a highly conformal and modulated boost to the target volume. (d) Spot map for the TB showing uniform grid-based spot placement over the target region. (e) DVHs for the CTV, ipsilateral lung, and skin.*

Modulation of spot monitor units (MUs) was used to enhance dose conformity by positioning the Bragg peaks on a high-density object (gold bar), located outside the patient's body

(Shown as a pink rectangle in Figure 1(a-c)). This setup enabled plan optimization analogous to a conventional proton beam approach in RayStation (Figure 1(b)). The different sizes of spots in Figure 1(d) reflect modulated MUs and an optimized TB plan. The MUs were optimized to achieve adequate target coverage while sparing OARS and reducing hot spots. The TB setup successfully delivered approximately 90% of the prescribed dose to the clinical target volume (CTV). As shown in Figure 1(c), the en face clinical beam provided a conformal background dose distribution, particularly to Anterior–Right (AR) and Posterior–Left (PL) regions. The remaining dose from this beam is configured with a 5 cm range shifter (RS) and constrained to a per-spot MU range of between 3 and 135 MUs. Figure 1(e) illustrates the resulting DVHs for the CTV, ipsilateral lung, and skin, showing acceptable target coverage and effective normal tissue sparing. The combined delivery approach enabled a treatment plan with dosimetric quality comparable to that of conventional Bragg peak-based PT.

The dose of 32 Gy in 5 fractions (6.4 Gy per fraction) was selected to achieve a high biologically effective dose (BED) and equivalent dose in 2 Gy fractions (EQD2), calculated to be approximately 100 Gy (BED) and 60 Gy (EQD2), respectively ($\alpha/\beta=3$), thus ensuring adequate biological effectiveness for gross disease control. Moreover, a fraction size greater than 5 Gy was chosen intentionally to maximize the potential FLASH radiotherapy effect[33]. This dose was selected after consideration of previously published five-fraction hypofractionated breast radiotherapy regimens in both the palliative and curative settings[34,35] and the promise of FLASH treatment to reduce normal tissue toxicity that otherwise limits total dose in 5-fraction regimens treating the whole breast.

## 2.2. FLASH models:

The dose and dose rate threshold required to trigger the FLASH effect range from 4-10 Gy and 40-100 Gy/s, respectively[29,36,37]. In this paper, we have used 4 Gy and 40 Gy/s as dose and dose rate thresholds. Unlike a passive scattering beam, the proton beam employs multiple spot scanning, delivering a dose to a given voxel at different time intervals. This temporal variation complicates the calculation of dose rate, which is essential for evaluating the FLASH effect. Various models have been developed, each employing a different interpretation for dose rate calculation[15,16,28,38]. In this study, we compare the FLASH effectiveness model proposed by Krieger et al. against a modified ADR as the means to determine FLASH effect[28].

### 2.2.1. FLASH effectiveness model (FEM):

In this model, the FLASH effect is triggered if a voxel receives a dose greater than 4 Gy within a time window (Δt = 200ms) at a dose rate greater than 40 Gy/s. Once FLASH is triggered, the effect remains persistent for another 200 ms, and the biological effect of 0.67 is reflected in the dose of the voxel during this time[32].

### 2.2.2. *Folkerts' Modified Average Dose Rate (ADR)*:

The ADR model used in this study is inspired by the voxel-specific model proposed by Folkerts et al., which incorporates inter-spot dead times and spot delivery times in pencil beam scanning. Details are provided in Supplementary Material (Section S2). This improved temporal resolution enables more biologically informed modeling of FLASH dose. Consistent with the FEM approach, the same adjustment was applied to voxels meeting both dose and dose-rate criteria.

### 2.3. Spot Scanning order:

In the Varian ProBeam system, the scan speed is direction dependent, as in the Y direction it is almost twice that in the X direction[39]. This anisotropy in scanning speed is further influenced by the inter-spot spacing, which plays a critical role in determining the effective delivery time and dose rate. The scanning magnets governing spot motion along the X and Y axes operate simultaneously, as such, the time required to transition between two consecutive spots is determined by the dominant axis and is calculated using the relation:

$$T_{scan} = \max\left(\frac{|\Delta x|}{v_x}, \frac{|\Delta y|}{v_y}\right) \quad (1)$$

Where $\Delta x$ and $\Delta y$ represent spatial displacements between adjacent spots in the X and Y directions, and $v_x$ and $v_y$ denote the corresponding scan speeds. The scan speed along the Y-axis plays an essential role in FLASH delivery, particularly since the default spot scanning pattern typically follows a zigzag trajectory in the Y-direction. In this study, the Y-axis scan speed was systematically varied between 10 mm/ms and 20 mm/ms to evaluate the impact of this parameter on the FLASH effect.

A tangential TB arrangement is employed to irradiate the chest-wall CTV. Due to this orientation, the temporal extent from the bottom to the top of each scanning column becomes relatively long. To maintain the dose delivery within the FLASH delivery window (200 ms from trigger initiation) for the FEM, the temporal distance on spot delivery becomes increasingly important. Therefore, optimizing the scanning sequence can potentially reduce overall dose delivery time and increase the possibility of achieving the FLASH effect on more voxels.

Three different scanning strategies for the evaluation of the FLASH effect were used:

1. Conventional Full-Field Scanning: A single TB was used to cover the entire CTV using the Bragg peak technique, with a hexagonal spot spacing of 6 mm to ensure comprehensive dose coverage, including the central region and edges of the target, Supplementary Figure S2 (Left Panel).

2. Split-Field Scanning: The single TB was divided into two separate TBs, each responsible for irradiating either the upper or lower half of the CTV. As shown in Supplementary Figure S2 (Right Panel). This approach effectively reduces the vertical travel distance within each column, thereby reducing temporal distance and better aligning with the FLASH window.

3. Optimized Scanning via Path Minimization: The spot delivery sequence was optimized by treating the problem as a variant of the Hamiltonian path problem. A well-known heuristic for solving instances of the traveling salesman problem (TSP) is the Genetic algorithm (GA), which is employed to generate near-optimal scanning paths of a single TB that reduce inter-spot travel time.

**2.3.1. Genetic Algorithm (GA):**

Among the various optimization algorithms, the GA demonstrates robustness and scalability, making it well-suited for handling many proton spots. Each combination of spots depicts a chromosome, and a spot is analogous to a gene in the evolution mechanism of the GA problem[40]. An initial population of candidates evolves over successive generations or iterations as they explore the randomness. This randomness is guided by stochastic operators such as crossover and mutation to emulate natural selection and genetic variability in each generation. The eventual goal of GA is to minimize the total delivery time, therefore maximizing the dose rate and enhancing the FLASH effect. Details regarding the mathematical formulation of the optimization problem and implementation of the workflow are provided in the supplementary material (Section S3).

**2.4. Patient study**

A dosimetric analysis of five left-sided post-mastectomy breast cancer patients was performed. All patients were planned with an ultra-hypofractionated proton dose of 32 Gy delivered in 5 fractions, with treatment plans ensuring that 95% of the prescribed dose covered the CTV. Each treatment plan consisted of two distinct beam sets: a TB and a clinical background beam. The single energy (249 MeV) TB, oriented tangentially (Left posterior oblique) in all cases to minimize exposure to OARs, delivered approximately 90% of the prescribed dose to the CTV. As the Bragg peak of the TB lies outside the patient's body, the dose was deposited primarily via the entrance (non-Bragg peak) region. To control superficial hot-spots, a 5 mm bolus was modeled in all cases, while the supplemented background beam achieves homogeneity. Along with a single TB plan, a split TB planning strategy was used in which the CTV was divided into two nearly equal sub-volumes, each irradiated by a separate TB of identical spot spacing (6 mm) and energy (249 MeV). Both approaches were further supported by the background beam.

The FLASH effect was evaluated for all plans using two frameworks, the FEM and Folkerts' ADR model. Quantitatively, each plan was analyzed and compared by metrics that included the percentage of FLASH-qualifying voxels within the CTV and skin volumes, with CTV Dmean analyzed as an additional parameter for biological impact. For effective model calculations, the beam current was fixed at 500 nA. In addition, CTV Dmean was analyzed as a measure of the biological impact by each model. Sensitivity analysis was also conducted by increasing the vertical scan speed from 10 mm/ms to 20 mm/ms to assess its influence on the outcomes of the FLASH models.

Optimization of the TB scanning spot patterns was performed using GA. Patient-specific parameters were used based on the number of TB spots and the spatial dimensions of each patient's CTV. Given the elongated geometry of the chest wall CTV, it posed a large-scale optimization problem involving numerous spots and extended scanning distances. The GA preserved fluence and CTV coverage, while optimizing the delivery order initialized with a population size of $\geq$3,000 and evolved over at least 5,000 generations. The dosimetric performance of GA-optimized sequences was compared against the default raster scanning pattern using CTV Dmean, skin Dmean, and skin D0.03cc.

## 4. Results:

The impact of vertical scan speed and dose rate model on FLASH effect was evaluated for both single and split-beam configurations. Figure 2 demonstrates the outcomes for Case 1 of this study. Increasing the y-direction scan speed from 10 mm/ms to 20 mm/ms led to a notable change in FLASH dose coverage across all models. In the single beam setting, the FEM yielded 13.2% voxels with FLASH dose in the CTV region at $V_y$=10 mm/ms, which increased to 30.5% at $V_y$=20 mm/ms. A similar trend was observed using the average dose rate model, with percentage FLASH effect of 37.1% and 67.9% for the respective scan speeds. For each scan speed, the dependence of the FLASH effect is evident, and the trend is similar across all studied cases.

In the split-beam configuration, further improvements were evident in lateral dose coverage, with peak FLASH effect reaching 92.1% using the FEM and 90.3% with the ADR model at $V_y$ = 20 mm/ms. The sensitivity of the FLASH effect to scan speed and model choice was also reflected in other dosimetric parameters. The smallest reduction in Dmean to the CTV was observed in Case 3, with a decrease of 1.07 Gy using FEM at $V_y$ = 10 mm/ms. In contrast, the greatest reduction in CTV Dmean was recorded in case 1 under the ADR model at $V_y$ = 20 mm/ms in the split-beam setting, amounting to 9.57 Gy. The trend is similar across all the cases. The percentage of voxels (averaged across all cases) that received FLASH dose in the CTV and skin is shown in Figure 3. The percentage increases with higher vertical scan speed ($V_y$=20 mm/ms), which is further enhanced by utilizing a split-beam (2TB) configuration. Across all scenarios, the ADR model consistently yields a higher proportion of FLASH than the FEM, with the highest values observed in the 2TB-ADR (Vy=20 mm/ms) scenario, reaching up to 94.84% in the CTV and 71.96% in the skin.

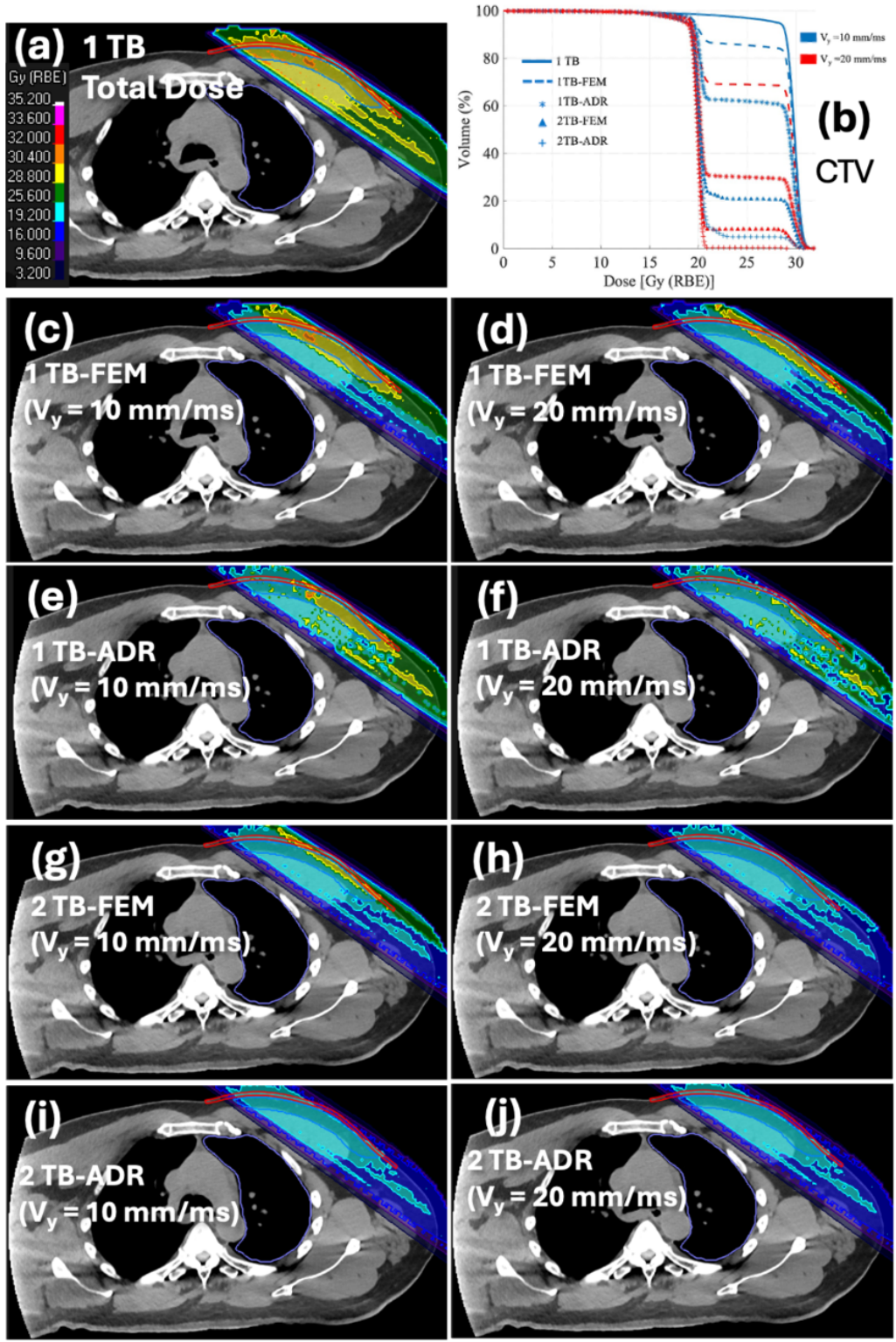

*Figure 2. (a) Dose distribution of a single transmission beam (TB). (b) Dose–volume histograms (DVHs) of CTV for all configurations, comparing FLASH effectiveness model (FEM) and average dose rate (ADR) models at scanning speeds of Vy = 10 and 20 mm/ms. (c–d) FLASH dose distributions computed using the FEM with scan speeds Vy =10 mm/ms and Vy =20 mm/ms, respectively. (e–f) Corresponding FLASH dose generated by using the average dose rate (ADR) model for the same scan speeds. (g–h) FLASH doses obtained via the FEM under split-beam configuration with Vy=10 mm/ms and Vy =20 mm/ms. (I–j) FLASH dose distributions calculated using the ADR model in split-beam configuration for Vy =10 mm/ms and Vy=20 mm/ms, respectively. In all cases, the horizontal scan speed is kept constant at Vx=10 mm/ms.*

Figure 4 illustrates the dose-volume histograms (DVHs) for the CTV and skin under different FLASH modeling approaches and scan speeds for case 1 of this study. Figure 4 (Left Panel) shows that the application of the FEM resulted in a reduction in both skin and CTV doses, with further dose sparing achieved through GA optimization of the scanning pattern under the $V_y$ = 10 mm/ms condition.

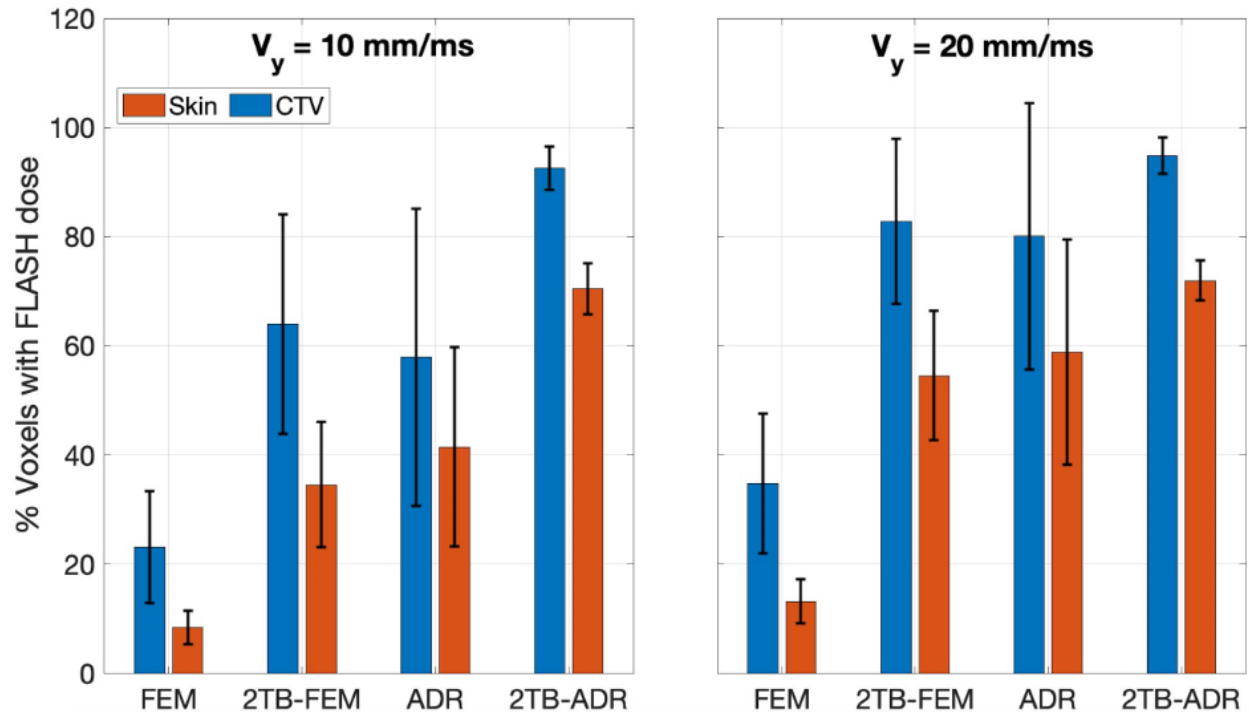

*Figure 3. Percentage of voxels (averaged over all cases) receiving FLASH dose in the CTV (blue) and skin (orange) for different FLASH models and beam delivery strategies. Results are grouped by vertical scan speed: Vy = 10 mm/ms (left) and Vy = 20 mm/ms (right). Each panel compares the FLASH Effectiveness Model (FEM) and the Average Dose Rate (ADR) model under single-beam (TB) and split-beam (2TB) configurations. Error bars denote the standard deviation across patients, representing inter-patient variability in FLASH coverage for each scenario.*

Similarly, the DVH shown in Figure 4 (Right Panel) reflects the superior skin sparing relative to the total dose across all models, with the most pronounced reductions observed in GA-optimized plans at $V_y$ = 10 mm/ms. A detailed, case-specific comparison of skin dosimetrics (Dmean and D0.03cc) with ADR and FEM based FLASH effects and the improvements achieved through scan pattern optimization is presented in Table S1. Across all cases, the ADR model was associated with a more substantial reduction in the Dmean to the CTV relative to FEM. Notably, the GA-optimized delivery under the ADR framework achieved FLASH effect comparable to those observed in the split-beam configuration using the default scan pattern.

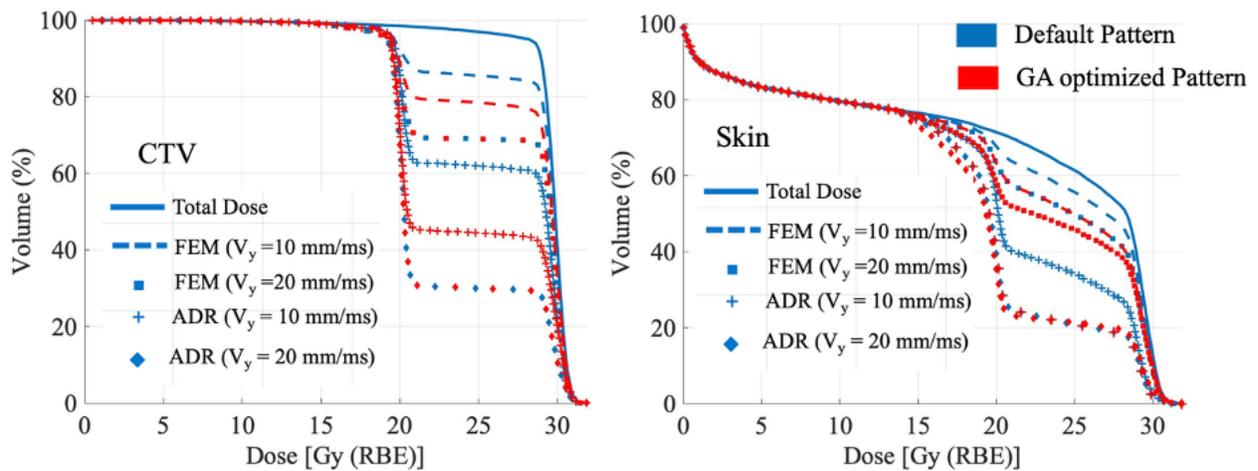

Figure 4. The DVHs receiving dose (Gy) in the CTV (left) and Skin (right) across multiple scenarios. The Total Dose curve represents the unmodified physical dose. FLASH dose was computed using either the FLASH Effectiveness Model (FEM) or the Average Dose Rate (ADR) model, with (red) and without (blue) GA optimization. Each configuration was tested under two vertical scanning speeds (Vy = 10 mm/ms and 20 mm/ms). GA-optimized sequences consistently shift DVHs downward, indicating a higher proportion of voxels experiencing FLASH-activated dose reduction, particularly in the skin.

A comprehensive comparison of all calculated parameter-dependent FLASH effects is presented in Figure 5. This figure illustrates pairwise differences in the Dmean to the CTV relative to the baseline TB configuration, enabling direct comparisons across configurations and models. The observed impact ranged from a lowest improvement of 2.27 Gy using the FEM at $V_y$ = 10 mm/ms, to a substantial reduction of 9.2 Gy achieved with the ADR model under a split-beam configuration.

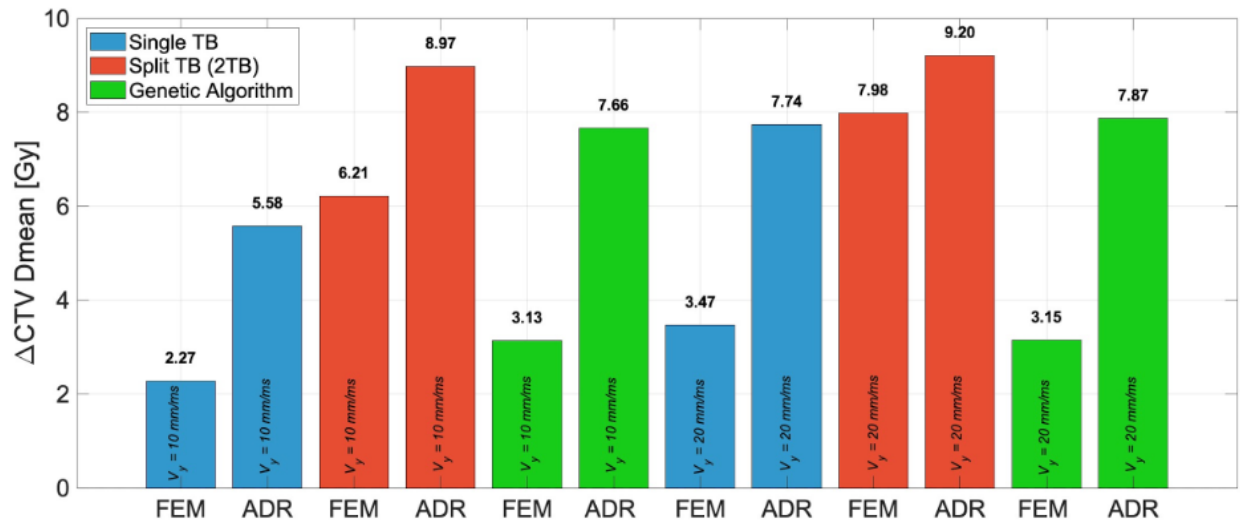

*Figure 5. Comparison of reductions in Dmean CTV (ΔCTV Dmean) under different FLASH models, beam configurations, and scan speeds (Averaged across all patients). The bar chart illustrates ΔCTV Dmean relative to the total physical dose for the FLASH effectiveness model (FEM) and the average dose rate model (ADR), across two vertical scan speeds, Vy =10 mm/ms and Vy = 20 mm/ms. Three planning strategies are compared: single transmission beam (Single TB, blue), split transmission beams (Split TB, red), and GA-based optimization (green). Results indicate that the ADR model consistently yields greater dose reductions than FEM for all configurations and scan speeds. Furthermore, GA-optimized plans demonstrate the highest reductions in CTV dose among all methods, with up to 8.97 Gy reduction at Vy=10 mm/ms under the ADR model.*

## 4. Discussion:

The FLASH effect due to proton TB for a large-field post-mastectomy chest-wall treatment strongly depends on the multiple interacting parameters, including vertical scan speed, dose rate evaluation model, and beam delivery configuration. The pursuit for a conclusive definition of the dose rate model is still underway. As this study shows, the choice of model can significantly influence anticipated outcomes.

The primary objective of FLASH effect is to spare normal tissues by delivering radiation at UHDR. Preclinical data suggest normal cells are less sensitive to UHDR due to hypoxia-induced radioresistance[41]. Tumor cells are believed to exhibit reduced susceptibility to the FLASH phenomenon and receive the full prescribed dose. This differential biological response suggests that achieving dose rates exceeding the 40 Gy/s threshold, particularly across all OARs, is essential. In contrast, the dose rate within the tumor or CTV may be of lesser concern. However, in the post-mastectomy cases, the bulk of malignant tissue is surgically removed, and the CTV predominantly comprises residual microscopic disease, if any disease at all. This makes the significance of dose rate within the CTV and proximal OARs more pronounced.

Furthermore, in the context of TB FLASH-RT, sparing of internal OARs, such as the ipsilateral lung and heart, can be achieved through breath-hold techniques, which reduce motion and displacement of thoracic structures during irradiation. However, the skin OAR presents a unique challenge. Due to its anatomical proximity and encapsulation of the CTV, the skin receives a similar dose rate distribution to that of the irradiated target region. This anatomical configuration, particularly in post-mastectomy treatment settings, limits the differential sparing of the skin achievable in conventional settings. Studies have employed the ray tracing method to identify individual spots that predominantly contribute to the dose or dose rate within specific OARs[31]. However, in this current setting, as seen from the BEV of tangential TB, the skin and CTV share a largely overlapping spatial profile, rendering spot-specific ray tracing less informative for differentiating contributions to dose rate between these adjacent structures.

All five patient cases included in this study underwent three distinct calculation scenarios to quantitatively evaluate the dependence of the FLASH effect on key parameters and models. Each of these scenarios is highlighted in Figure 5 with blue, green, and red color bars. In each scenario, a pair of calculations reflects the vertical scan speed ($V_y$) varied from 10 mm/ms to 20 mm/ms, and both the ADR model and the FEM were employed to analyze FLASH dose distributions. Across all cases, increasing the vertical scan speed consistently led to an enhanced FLASH effect to both the CTV and the skin. Additionally, the most significant decrease in the Dmean of CTV was observed for the split beam setting. The split beam configuration showed improvements by reducing the vertical scan and decreasing the temporal distances, particularly to large or elongated target geometries, as shown in Figure S2.

Regarding the model selection, the ADR model consistently demonstrated a higher FLASH effect compared to the FEM across all scenarios, particularly within the CTV and skin. This trend is quantitatively illustrated in Figure 3, which displays the percentage of voxels receiving FLASH dose across both regions for each model and delivery configuration. The lowest FLASH coverage was observed under the single TB configuration at a vertical scan speed ($V_y$) of 10 mm/ms, yielding approximately 40% FLASH activation in CTV voxels. However, due to the default spot scanning sequence, the FLASH effect under this configuration was spatially biased, predominantly concentrated on the inner surface of the CTV, as visualized in the 2D dose distribution maps shown in Figure 2(c–e). To achieve a more uniform FLASH effect across the entire CTV, configurations employing a split-beam setup with the ADR model proved more effective. In particular, the highest apparent FLASH coverage, reaching up to 95%, was obtained using the ADR model. An increase from 40% to 95% underscores the profound sensitivity of FLASH outcome to variations in scan speed, beam configuration, and model selection.

The difference in FLASH effect arises from the interpretation of temporal dose dynamics in the two models. The ADR model accounts for the cumulative temporal structure of spot-scanning by calculating voxel-specific dose rates over an effective irradiation interval, offering a time-resolved representation of dose accumulation. This approach reflects the physical conditions required to trigger radiolytic oxygen depletion mechanisms of the FLASH effect. In contrast, the FEM adopts a more discrete framework by assessing dose and dose rate within short temporal windows. If both the dose and dose rate exceed threshold values within any of these brief intervals, FLASH is considered to be triggered. Another unique characteristic of FEM is the persistence time parameter, which guarantees that the FLASH effect may persist biologically even after the initial dose-rate criteria are no longer met.

The FLASH effect, with its dependence on the dose rate, is intrinsically linked to the overall beam scan time. Improvements in FLASH delivery can be achieved through optimization of the proton spot scanning pattern. This scan pattern optimization problem is formulated analogously to the Travelling Salesman Problem (TSP) and optimized using a GA. Across all studied cases, this GA optimization strategy yielded improvements in both FLASH dose metrics and dose reduction to OARs. Notably, under the $V_y$ = 10 mm/ms condition, the optimizer exhibited greater flexibility in reordering spot sequences, resulting in a noticeable reduction in scan time and enhanced FLASH effect. In contrast, under the $V_y$ = 20 mm/ms setting, the default scanning pattern (shown in Figure S2) has already approximated the minimum achievable scan time. Due to directional constraints imposed by the higher y-axis scan speed, the optimized scan pattern closely resembled the default configuration, resulting in a comparatively similar FLASH effect.

Clinically, the FLASH effect in the skin highlights the potential for significant reductions in acute and late toxicity as reported in Figure 4 (Skin panel) and Table S1. As reflected in Table S1, there were significant reductions in the mean skin dose in all cases with both models. Importantly, as seen in Figure 4, optimal skin dose sparing was achieved through optimization of the spot delivery sequence, without modifying beam energy, spot spacing, or compromising target coverage. It should be noted that while improvements were seen in the FLASH-effect dose adjusted mean skin doses, the maximum skin doses remain largely unaffected across the various calculation scenarios, likely due to the anatomical configuration in which the skin closely envelops the CTV.

The dose and dose-rate thresholds used in this study, 4 Gy and 40 Gy/s, respectively, were held constant across all analyses. However, it is important to note that these thresholds vary across the literature, and different values could alter the reported extent of FLASH dose coverage. A key parameter influencing dose rate is the spot delivery time, which is defined by the ratio of spot MUs to beam fluence. Both spot MUs and fluence are dependent on vendor-specific system characteristics and delivery mechanisms; hence, they will show variability across different clinical platforms. A nozzle current of 500 nA was used for fluence calculations in this study. However, this value is also system-dependent and may vary depending on the specific accelerator used in the FLASH study. The fixed parameters utilized here, including current, dose thresholds, and system configuration, should be interpreted within the context of the Varian ProBeam system, on which this study is based. These parameters significantly influence FLASH evaluations and may limit the generalizability of results across other delivery platforms. Moreover, while GA optimization was applied at the treatment planning level, its downstream implications on machine delivery efficiency, interplay effects, and quality assurance (QA) workflows remain to be investigated. The current TB based delivery paradigm assumes breath-hold conditions to minimize intrafraction motion, however, such motion control may not be consistently achievable across all clinical scenarios, which could impact the reproducibility and robustness of FLASH treatments.

In summary, this study evaluated how scanning pattern, scan speed, and dose-rate modeling affect FLASH proton radiotherapy plans for post-mastectomy chest wall treatment. GA-based path optimization reduced scan time and improved FLASH dose coverage without compromising target coverage. Further clinical studies are needed to confirm these dosimetric findings.

# 6. References:

1. Surveillance E, and End Results (SEER) Program, National Cancer Institute. Cancer Stat Facts: Female Breast Cancer. Internet. National Cancer Institute. 2025. 2025. https://seer.cancer.gov/statfacts/html/breast.html
2. Schreuder K, Middelburg JG, Aarts MJ, et al. An actualised population‐based study on the use of radiotherapy in breast cancer patients in the Netherlands. *The Breast Journal*. 2019;25(5):942-947.
3. Gao RW, Mullikin TC, Aziz KA, et al. Postmastectomy intensity modulated proton therapy: 5-year oncologic and patient-reported outcomes. *International Journal of Radiation Oncology* Biology* Physics*. 2023;117(4):846-856.
4. Marlière S, Vautrin E, Saunier C, Chaikh A, Gabelle-Flandin I. Radiation-related heart toxicity: Update in women. 2016:411-419.
5. Darby SC, Ewertz M, McGale P, et al. Risk of ischemic heart disease in women after radiotherapy for breast cancer. *New England Journal of Medicine*. 2013;368(11):987-998.
6. Taylor C, Correa C, Duane FK, et al. Estimating the risks of breast cancer radiotherapy: evidence from modern radiation doses to the lungs and heart and from previous randomized trials. *Journal of Clinical Oncology*. 2017;35(15):1641-1649.
7. Group EBCTC. Effect of radiotherapy after breast-conserving surgery on 10-year recurrence and 15-year breast cancer death: meta-analysis of individual patient data for 10 801 women in 17 randomised trials. *The Lancet*. 2011;378(9804):1707-1716.
8. Taylor CW, Wang Z, Macaulay E, Jagsi R, Duane F, Darby SC. Exposure of the heart in breast cancer radiation therapy: a systematic review of heart doses published during 2003 to 2013. *International Journal of Radiation Oncology* Biology* Physics*. 2015;93(4):845-853.
9. Ranger A, Dunlop A, Hutchinson K, et al. A dosimetric comparison of breast radiotherapy techniques to treat locoregional lymph nodes including the internal mammary chain. *Clinical Oncology*. 2018;30(6):346-353.
10. Bradley JA, Dagan R, Ho MW, et al. Initial report of a prospective dosimetric and clinical feasibility trial demonstrates the potential of protons to increase the therapeutic ratio in breast cancer compared with photons. *International Journal of Radiation Oncology* Biology* Physics*. 2016;95(1):411-421.
11. Mast ME, Vredeveld EJ, Credoe HM, et al. Whole breast proton irradiation for maximal reduction of heart dose in breast cancer patients. *Breast cancer research and treatment*. 2014;148(1):33-39.
12. Jimenez RB, Goma C, Nyamwanda J, et al. Intensity modulated proton therapy for postmastectomy radiation of bilateral implant reconstructed breasts: a treatment planning study. *Radiotherapy and Oncology*. 2013;107(2):213-217.
13. Ares C, Khan S, MacArtain AM, et al. Postoperative proton radiotherapy for localized and locoregional breast cancer: potential for clinically relevant improvements? *International Journal of Radiation Oncology* Biology* Physics*. 2010;76(3):685-697.
14. Mou B, Beltran CJ, Park SS, Olivier KR, Furutani KM. Feasibility of proton transmission-beam stereotactic ablative radiotherapy versus photon stereotactic ablative radiotherapy for lung tumors: A dosimetric and feasibility study. *PLoS One*. 2014;9(6):e98621.

15. Van De Water S, Safai S, Schippers JM, Weber DC, Lomax AJ. Towards FLASH proton therapy: the impact of treatment planning and machine characteristics on achievable dose rates. *Acta oncologica*. 2019;58(10):1463-1469.
16. van Marlen P, Dahele M, Folkerts M, Abel E, Slotman BJ, Verbakel WF. Bringing FLASH to the clinic: treatment planning considerations for ultrahigh dose-rate proton beams. *International Journal of Radiation Oncology* Biology* Physics*. 2020;106(3):621-629.
17. van Marlen P, Dahele M, Folkerts M, Abel E, Slotman BJ, Verbakel W. Ultra-high dose rate transmission beam proton therapy for conventionally fractionated head and neck cancer: Treatment planning and dose rate distributions. *Cancers*. 2021;13(8):1859.
18. Vozenin M-C, De Fornel P, Petersson K, et al. The advantage of FLASH radiotherapy confirmed in mini-pig and cat-cancer patients. *Clinical Cancer Research*. 2019;25(1):35-42.
19. Favaudon V, Caplier L, Monceau V, et al. Ultrahigh dose-rate FLASH irradiation increases the differential response between normal and tumor tissue in mice. *Science translational medicine*. 2014;6(245):245ra93-245ra93.
20. Montay-Gruel P, Petersson K, Jaccard M, et al. Irradiation in a flash: Unique sparing of memory in mice after whole brain irradiation with dose rates above 100 Gy/s. *Radiotherapy and Oncology*. 2017;124(3):365-369.
21. Simmons DA, Lartey FM, Schüler E, et al. Reduced cognitive deficits after FLASH irradiation of whole mouse brain are associated with less hippocampal dendritic spine loss and neuroinflammation. *Radiotherapy and Oncology*. 2019;139:4-10.
22. van Marlen P, van de Water S, Dahele M, Slotman BJ, Verbakel WF. Single ultra-high dose rate proton transmission beam for whole breast FLASH-irradiation: quantification of FLASH-dose and relation with beam parameters. *Cancers*. 2023;15(9):2579.
23. Ma C, Yang X, Setianegara J, et al. Feasibility study of modularized pin ridge filter implementation in proton FLASH planning for liver stereotactic ablative body radiotherapy. *Physics in Medicine & Biology*. 2024;69(24):245001.
24. AJ Zafar, Yang X, Diamond Z, et al. An adaptive proton FLASH therapy using modularized pin ridge filter. *Medical Physics*. 2025;52(9):e18109.
25. Forster T, Köhler CVK, Debus J, Hörner-Rieber J. Accelerated partial breast irradiation: a new standard of care? *Breast Care*. 2020;15(2):136-147.
26. Folkerts MM, Abel E, Busold S, Perez JR, Krishnamurthi V, Ling CC. A framework for defining FLASH dose rate for pencil beam scanning. *Medical physics*. 2020;47(12):6396-6404.
27. Pin A, Tibi D, Hotoiu L, et al. Pencil beam proton flash therapy, field size limit with ConformalFLASH. *Physica Medica: European Journal of Medical Physics*. 2022;94:S66-S67.
28. Krieger M, Van De Water S, Folkerts MM, et al. A quantitative FLASH effectiveness model to reveal potentials and pitfalls of high dose rate proton therapy. *Medical physics*. 2022;49(3):2026-2038.
29. Wilson JD, Hammond EM, Higgins GS, Petersson K. Ultra-high dose rate (FLASH) radiotherapy: Silver bullet or fool's gold? *Frontiers in oncology*. 2020;9:1563.
30. Petersson K, Adrian G, Butterworth K, McMahon SJ. A quantitative analysis of the role of oxygen tension in FLASH radiation therapy. *International Journal of Radiation Oncology* Biology* Physics*. 2020;107(3):539-547.

31. Zhao X, Huang S, Lin H, et al. A novel dose rate optimization method to maximize ultrahigh-dose-rate coverage of critical organs at risk without compromising dosimetry metrics in proton pencil beam scanning FLASH radiation therapy. *International Journal of Radiation Oncology* Biology* Physics*. 2024;120(4):1181-1191.
32. Wase V, Widenfalk O, Nilsson R, Fälth C, Fredriksson A. Fast spot order optimization to increase dose rates in scanned particle therapy FLASH treatments. *Physics in Medicine & Biology*. 2025;70(2):025017.
33. Böhlen TT, Germond JF, Bourhis J, Bailat C, Bochud F, Moeckli R. The minimal FLASH sparing effect needed to compensate the increase of radiobiological damage due to hypofractionation for late‐reacting tissues. *Medical physics*. 2022;49(12):7672-7682.
34. Chatterjee S, Chakrabarty S, Santosham R, et al. Alleviating morbidity from locally advanced breast cancer using a practical and short radiation therapy regimen: results of the HYPORT palliative studies. *International Journal of Radiation Oncology* Biology* Physics*. 2023;116(5):1033-1042.
35. Brunt AM, Haviland JS, Wheatley DA, et al. Hypofractionated breast radiotherapy for 1 week versus 3 weeks (FAST-Forward): 5-year efficacy and late normal tissue effects results from a multicentre, non-inferiority, randomised, phase 3 trial. *The Lancet*. 2020;395(10237):1613-1626.
36. Chabi S, Van To TH, Leavitt R, et al. Ultra-high-dose-rate FLASH and conventional-dose-rate irradiation differentially affect human acute lymphoblastic leukemia and normal hematopoiesis. *International Journal of Radiation Oncology* Biology* Physics*. 2021;109(3):819-829.
37. Montay-Gruel P, Bouchet A, Jaccard M, et al. X-rays can trigger the FLASH effect: Ultra-high dose-rate synchrotron light source prevents normal brain injury after whole brain irradiation in mice. *Radiotherapy and Oncology*. 2018;129(3):582-588.
38. Schwarz M, Traneus E, Safai S, Kolano A, van de Water S. Treatment planning for Flash radiotherapy: General aspects and applications to proton beams. *Medical Physics*. 2022;49(4):2861-2874.
39. Tsai P, Yang Y, Wu M, et al. A comprehensive pre‐clinical treatment quality assurance program using unique spot patterns for proton pencil beam scanning FLASH radiotherapy. *Journal of Applied Clinical Medical Physics*. 2024;25(8):e14400.
40. Rao A, Hegde SK, Rao A, et al. Literature survey on travelling salesman problem using genetic algorithms. *International Journal of Advanced Research in Eduation Technology (IJARET)*. 2015;2(1):42.
41. Spitz DR, Buettner GR, Petronek MS, et al. An integrated physico-chemical approach for explaining the differential impact of FLASH versus conventional dose rate irradiation on cancer and normal tissue responses. *Radiotherapy and oncology*. 2019;139:23-27.
42. Vozenin M-C, Hendry JH, Limoli C. Biological benefits of ultra-high dose rate FLASH radiotherapy: sleeping beauty awoken. *Clinical oncology*. 2019;31(7):407-415.
43. Kang M, Wei S, Choi JI, Simone CB, Lin H. Quantitative assessment of 3D dose rate for proton pencil beam scanning FLASH radiotherapy and its application for lung hypofractionation treatment planning. *Cancers*. 2021;13(14):3549.

**Supplementary Material:**

**S1. FLASH Effectiveness Model (FEM):**

In this model, the determination of flash effect is based on the calculation of dose ($D_0$) delivered to each voxel within a specific window of time ($\Delta t$). If $D_0$ exceeds the dose threshold values (4 Gy) and the average dose rate ($\dot{D}_0$) in that instant meets the threshold value of 40 Gy/s, the model triggers the FLASH trigger in the healthy tissues. The activation of the FLASH in the FLASH-triggered window depends on the temporal duration of dose delivery within the interval $\Delta t$. Furthermore, once triggered, the FLASH condition remains active for an additional persistence time of 200 ms. During the active FLASH phase, the biological effect of the dose delivered to each voxel of healthy tissue is reduced. To account for this diminished toxicity, a dose adjustment factor (<1) of 0.67 is applied. The adjustment factor reflects a 33% reduction in the dose delivered to each voxel during FLASH triggered window. This effect would typically not be applied within tumor tissue. Although in the case of PMRT, the CTV is mostly normal tissue with perhaps microscopic tumor cells within.

In this study, the dose adjustment factor was likewise applied to the CTV region, although presuming the tumor cells within would not be affected by this.

**S2. Average Dose Rate (ADR) Model:**

The ADR model determines whether flash effects conditions are met by calculating the total dose received by that voxel divided by its effective time.

The effective dose delivery time is determined by identifying the time interval during which the voxel acquires the bulk of its dose. The interval begins at $t_i$ when the voxel's cumulative dose exceeds a pre-defined threshold $d^*$, and ends at $t_f$ when the remaining dose $(D_T - d^*)$ is delivered. This thresholding approach accounts for the biological relevance of rapid dose accumulation, which is thought to be essential for triggering the FLASH effect through mechanisms such as radiolytic oxygen depletion[42]. The dose rate for each voxel is given by:

$$\dot{D} = \frac{D_T - 2d^*}{t_f - t_i} \quad (2)$$

Here $D_T$ is the total dose accumulated by the voxel. The dose threshold used in this work is 0.1 Gy[26].

This conventional ADR approach proposed by Folkerts et al. estimates the dose rate using a threshold-based time window derived only from high-dose spot contributions[43]. This modified approach used in this study includes a voxel-specific delivery time, which incorporates all contributing spots with explicit scan and spot delivery timing. This enhanced temporal resolution enables biologically informed FLASH modeling. Like Krieger's suggestion, a fixed adjustment factor is then applied to voxels meeting both dose and dose-rate thresholds to describe the FLASH-effective dose per voxel.

**S3. Genetic Algorithm (GA):**

Each spot within a single energy layer of 249 MeV is initially assigned an index, referred to as a scanning pattern number. These indices serve as genes in the GA, where each scanning sequence corresponds to a chromosome representing a potential solution[40]. An initial population of size $P$ is generated, with each individual representing a distinct scanning pattern. Through each generation, genes (i.e., indexed spot positions) are stochastically recombined to form new chromosomes, while simulating biological crossover and mutation processes. This iterative procedure continues over successive generations, each aimed at improving the scanning path, until either a predetermined number of generations is reached or the optimization objectives, such as minimized scanning time or maximized FLASH dose rate, are satisfied.

To optimize the dose rate ($\Delta D/\Delta t$) for maximum FLASH effect, the time taken to deliver the dose at any particular voxel should be minimized. A set of $N$ spots with non-zero MUs in a single layer is represented by $S = \{s_1, s_2, s_3, \ldots, s_N\}$, where each $s_i$ has 2-dimensional coordinates $r_i = \mathbb{R}^d$. A scanning pattern (or chromosome) in the GA is a permutation $\pi = \{\pi(1), \pi(2), \pi(3), \ldots, \pi(N)\}$ of the spot indices representing the order in which the scanning occurs. The objective function $f(\pi)$, representing the total scanning time for a given sequence $\pi$ is defined as:

$$f(\pi) = \frac{1}{v} \sum_{i=1}^{N-1} \|r_{\pi(i+1)} - r_{\pi(i)}\|$$

Here $r_{\pi(i)} = (x_{\pi(i)}, y_{\pi(i)})$ are 2D coordinates of spot $\pi$, and $v$ is the scan velocity. This objective function is minimized during the GA optimization. Since $v$ is constant, the optimization effectively focuses on minimizing the total path length

$$\min_{\pi \in \mathcal{P}(S)} f(\pi)$$

$\mathcal{P}(S)$ denotes the set of all permutations of $S$.

Additionally, the initial population was initiated using both a randomly ordered individual and a nearest-neighbor heuristic-based path to accelerate convergence and have a jump start on better scanning patterns from the beginning. Furthermore, individual optimization was parallelized using Python's multiprocessing module to improve computational efficiency, since all cases had a large spot set. A complete schematic diagram of the GA optimization process is shown in Figure S1.

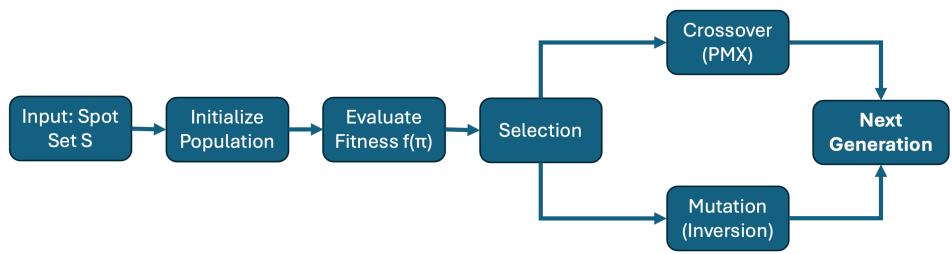

Figure S1. Genetic Algorithm Workflow for Spot Scanning Optimization.

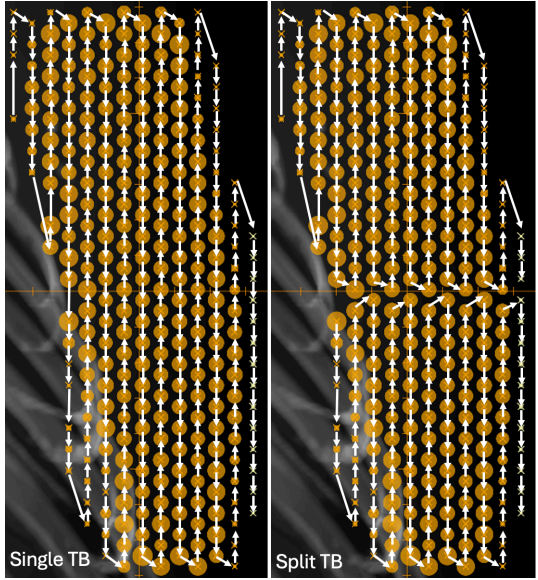

Figure S2. Spot scanning pattern of single energy TB showing the default path followed by the beam delivery setup, and Split beam scanning pattern.

Table S1. Summary of scan time, target coverage, and skin dose metrics across five cases for TB proton FLASH plans with and without genetic algorithm (GA) optimization under both FLASH Effectiveness Model (FEM) and Average Dose Rate (ADR) Model. The table reports original and GA-optimized scan times, average dose to CTV (Dmean CTV), maximum dose to 0.03 cc of skin (Skin0.03cc), and mean skin dose (Dmean Skin) for each case. TB values represent the total physical dose baseline, while "Default" and "GA-Optimized" columns reflect dose values calculated under FEM and ADR models, both before and after GA optimization. GA-optimized plans consistently reduced scan time while lowering skin dose metrics (Dmean) and CTV dose, while slightly reducing (Skin0.03cc). The ADR model generally calculates lower dose values than the FEM, particularly for skin metrics, showing the sensitivity of the choice of parameters and model in FLASH planning.

| | Optimize Scan time (ms) | | Dmean CTV (Gy) | | | Skin0.03cc (Gy) | | | Dmean Skin (Gy) | | |
|---|---|---|---|---|---|---|---|---|---|---|---|
| Case | Origional | GA-Optimized FEM / ADR | TB | Default FEM / ADR | GA-Optimized FEM / ADR | TB | Default FEM / ADR | GA-Optimized FEM / ADR | TB | Default FEM / ADR | GA-Optimized FEM / ADR |
| 1 | 364.39 | 340.33 / 340.33 | 29.48 | 28.33 / 25.97 | 27.63 / 24.26 | 31.53 | 31.01 / 31.41 | 30.88 / 31.41 | 31.53 | 21.29 / 19.02 | 20.19 / 17.47 |
| 2 | 333.39 | 327.67 / 311.09 | 29.08 | 26.19 / 20.33 | 25.45 / 19.64 | 31.68 | 31.68 / 30.01 | 30.75 / 30.36 | 31.68 | 21.92 / 17.06 | 19.14 / 16.35 |
| 3 | 523.6 | 502.93 / 490.45 | 29.25 | 28.17 / 27.14 | 27.35 / 22.94 | 31.57 | 31.5 / 31.5 | 31.57 / 31.5 | 31.57 | 22.62 / 21.57 | 21.18 / 18.72 |
| 4 | 421.98 | 414.3 / 396.83 | 29.33 | 26.47 / 22.63 | 25.95 / 20.5 | 30.72 | 30.57 / 30.48 | 30.63 / 30.43 | 30.72 | 23.15 / 19.81 | 22.41 / 17.31 |
| 5 | 361.27 | 345.99 / 343.95 | 28.88 | 25.49 / 21.98 | 23.99 / 20.4 | 30.74 | 30.74 / 30.59 | 30.74 / 30.59 | 30.74 | 22.96 / 18.52 | 21.78 / 17.26 |